\documentclass[journal,draft,onecolumn,12pt,twoside]{IEEEtran}
\usepackage{amsmath, amssymb, stmaryrd} 
\usepackage[final]{graphicx}
\usepackage{wrapfig}
\usepackage{lscape}
\usepackage{rotating}
\usepackage{amsmath}
\usepackage{color}
\usepackage{hyperref}
\usepackage{multirow}
\usepackage{fancyhdr}
\usepackage[normal,bf,up]{caption2}
\usepackage{subfigure}
\usepackage[section]{placeins}
\usepackage{algorithm} 
\usepackage[algo2e]{algorithm2e} 
\usepackage{mathabx}
\usepackage{xcolor}
\usepackage{algcompatible}
\usepackage{float}
\newfloat{algorithm}{t}{lop}
\newcommand{\stxt}[1]{\ensuremath{_{\text{#1}}}}

\ifCLASSINFOpdf
\else
\fi
\begin{document}
%
\title{Unsourced Random Access with Threshold$-$Based Feedback}
%
%
%

\author{Murwan Bashir, Ehsan Nassaji, Dmitri Truhachev, Alireza Bayesteh, Monirosharieh Vameghestahbanati}
\maketitle

\begin{abstract}
In this paper we focus on a feedback mechanism for unsourced random access (URA) communications. We propose an algorithm to construct feedback packets broadcasted to the users by the base station (BS) as well as the feedback packet format that allows the users to estimate their channels and infer positive or negative feedback based on the presented thresholding algorithm. We demonstrate that the proposed feedback imposes a much smaller complexity burden on the users compared to the feedback that positively acknowledges all successful or negatively acknowledges all undecoded users. We also show that the proposed feedback technique can lead to a substantial reduction in the packet error rates and signal-to-noise ratios (SNR)s required to support various numbers of active users in the system.
\end{abstract}


%
\IEEEpeerreviewmaketitle

\section{Introduction}
\label{LitSurvey}

Unsourced random access (URA) is a grant-free multi-user communication setting where a multitude of active users  transmit their messages to a sink point in a sporadic manner, without a mechanism to identify the transmitting users~\cite{polyanskiy2017perspective}. The time-varying set of active users is a subset of a much larger pool of users that may eventually communicate. In URA applications, the unique identifiers (ID)s of the active users are unimportant and the receiver is only interested in the message content itself. Such setting presents itself naturally in various sensor network applications, where the knowledge of the mapping between the messages and user IDs does not add extra value. Absence of user IDs is especially important for the case of short packet transmission, since inclusion of a user ID into the message will significantly reduce the data portion of the packet. Moreover, the complexity of a multiple user detection receiver that pre-allocates resources based on pre-defined IDs of all possible users can be extremely high for a large pool of potential users. 

The ID-free access strategy also reflects itself in the measurement of the URA performance. While the traditional multi-access communications targets high reliability across all users, the goal of the URA is to ensure that the majority of the transmitted users' messages are recovered. Hence, the per user probability of error (PUPE) has been proposed in~\cite{polyanskiy2017perspective} to measure URA performance reflecting the focus on the number of successfully delivered messages, rather than on delivering all messages error-free. 



Multiple URA systems proposed in the literature attempt to approach the information-theoretic lower bound derived in~\cite{polyanskiy2017perspective} in terms of the minimum SNRs required to support a given number of active users at a specific level of PUPE. Because of the sparse nature of the user activity in URA, where only a small number of users are active at any time slot, design of access methods that enable compressed sensing (CS) detection was among the first attempts of URA implementations. For example, in \cite{amalladinne2020coded,amalladinne2020approximate} the transmitted packet is partitioned into sub-blocks that are encoded by a code constructing a binary tree. The receiver first utilizes CS detection algorithms to recover the sets of active user message partitions for each sub-block. Following that, the tree decoders reconstruct the complete user messages from these sets. 

Another large class of URA systems, capable to support high numbers of active users, is based on the preamble-payload format~\cite{fengler2019sparcs}\cite{nassaji2021unsourced}. 
In the preamble-payload format the packet is sub-divided into a small preamble and a larger payload. The preamble carries a part of the message data and, at the same time, enables channel access by acting as a temporary user ID. CS detection algorithms are typically used to recover the preamble part. The payload is usually constructed via preamble-dependent encoding and modulation through permutation and scrambling. At the receiver, the payloads are retrieved using multi-user detection algorithms with some form of interference cancellation and error-correction decoding. 


In general, a URA system allows for packet transmission without enforcing reliability as an aspect of the system design. However, informing active users about the status of their received messages is beneficial in terms of stopping frequent re-transmissions. Moreover, implementation of a feedback to enhance reliability enables the design of re-transmission protocols that further reduce the required system SNR compared to the feedback-free URA as we demonstrate in this paper. We also show that the feedback can help saving power at the active user nodes which are often low-power battery-operated devices. 

The classic grant-free communications follows two main approaches to ensure low error probability of the packet reception. The first approach targets reduction of the packet collision probability by simple packet repetition~\cite{huda2019uplink} or linear network coding~\cite{choi2021network}. However, in this approach, the receiver is only benefiting from repetition diversity or the imposed error-correction, and any knowledge about the successful delivery of a packet is not communicated back to the active users. The second approach introduces hybrid automatic repeat request (HARQ) feedback protocol, where the receiver broadcasts the decoding status to the transmitters. It has been shown that the latter outperforms the repetition-like approaches in terms of reliability~\cite{ostman2018low} at the cost of some additional signaling. For example, in a technique proposed in \cite{choi2020throughput}\cite{zhang2020hybrid} the BS sends a negative acknowledgment (NACK) to colliding users in an attempt to stop simultaneous  re-transmissions.  Similarly, the status of packet decoding can be fed back to assist in scheduling and collision mitigation as shown in \cite{ren2021random} \cite{lee2016successive}.
In~\cite{yang2016compressed}, for the case of low number of active users,  positive feedback is broadcasted to the successful active users. The users employ a message-passing algorithm to detect their signatures in the feedback message and decide on the type of the received feedback (ACK or NACK). If active users can obtain channel state information, power-based threshold access can be used to enhance packet reliability at the receiver~\cite{tang2005opportunistic}. 

In asynchronous grant-free transmission settings feedback is utilized to inform candidate active users waiting for re-transmission about vacant channel slots~\cite{jung2012asynchronous}. Another utilization of feedback, focused on resource allocation, is presented in~\cite{facenda2020efficient}, where the feedback is used in a grant-based setting to avoid collisions and allow for user scheduling. In \cite{facenda2020efficient}, the authors advocate that their proposed alternative system can be used to perform the same function as the URA system presented in \cite{polyanskiy2017perspective}. In \cite{zhang2020hybrid} a grant-free access system  with feedback is proposed to resolve collisions using orthogonal pilots during re-transmission. A similar feedback strategy is employed in \cite{ding2020triangular} to prevent user collisions in case active users have unequal data packets sizes.

In URA setting, the feedback has been used very recently to enhance system reliability. In \cite{newFeedback} a targeted positive acknowledgement using a beam-forming system is proposed and it is shown that the length of feedback messages needs to scale linearly with the targeted number of users.  In \cite{singh2020minimum}\cite{kang2020minimum} the upper bound on the number of bits required for feedback to ensure collision-free transmission is provided. 

In addition to collision mitigation and enhancing reliability, feedback is employed to enable other aspects of networking such as security. For example, in \cite{kotaba2021identify}, each user utilizes a secret key that is known to the receiver to generate a message authentication code for message verification at the receiver. Feedback can also be utilized to minimize the age of information (AoI) \cite{kaul2011minimizing} which measures the freshness of information packets in time-sensitive applications. In \cite{chen2020age}, feedback is utilized to aid in the operation of a distributed transmission policy that minimizes the overall AoI. 




In this paper, we study a generic problem of setting up feedback in a URA system and identify the advantages the feedback can bring to URA. 
We start with two baseline feedback mechanisms, the ``positive-only feedback'' in which the BS sends successful acknowledgement to all successful users, and the ``negative-only feedback'' where the BS sends successful acknowledgements to all detected but undecoded users. 
While URA systems do not utilize fixed user IDs, focusing on preamble-payload URA allows us to work with feedback messages formed as a superposition of a subset of all active user preambles broadcasted back to the user pool. We then propose a feedback format in which the users can estimate their channels and use thresholding to infer a success or failure of their packet on the feed-forward link or receive a personalized feedback message. We further develop this idea to construct a system where most of the users can save power by processing only a fraction of the feedback message. In order to describe the system mathematically and quantify the gains of the proposed feedback mechanism numerically, we utilize the preamble payload format presented in \cite{nassaji2021unsourced}\cite{truhachev2020low}. We note, however, that the proposed techniques are applicable to any preamble-payload URA format.

Our contributions can be summarized as follows: 
\begin{itemize}
    \item We propose a general threshold-based feedback framework for URA systems. 
    We demonstrate that the proposed approach improves reliability while conserving the overall system resources compared to the positive-only or negative-only baseline feedback techniques.  
    \item We demonstrate that the proposed feedback results in a significant reduction in the SNRs required to support high numbers of active users. We show that the system with feedback and a single re-transmission opportunity can provide a significant advantage over a feed-forward only URA. 
    \item We present a double-threshold feedback system to enable further complexity reduction on the active user side by allowing partial processing of the feedback message. 
\end{itemize}

The paper is organized as follows. Section~\ref{SystemModel} presents the system model and the structure of the feed-froward link at the transmitter and receiver. In Section~\ref{sec:FB_link} we present two baseline feedback systems of positive and negative feedback before presenting the proposed single and double threshold-based feedback approaches. We then propose the utilization of partial signatures and the feedback detection mechanism at the active users side. 
Section~\ref{sec:NumericalResults} presents the numerical results, and  Section~\ref{sec:Con} concludes the paper.

\section{System Model}
\label{SystemModel}

\subsection{Feed-Forward Link}
\label{se:FFLink}

In the preamble$-$payload URA approach~\cite{fengler2019sparcs,nassaji2021unsourced,truhachev2020low,pradhan2019joint}, a packet of an active user consists of two sub-sections: the preamble and the payload. Generally, the preamble simultaneously serves the purpose of temporary user identification\footnote{The user is identified by the preamble selected at random from a common preamble pool, contrary to the permanent user ID used in conventional grant-free systems. Preamble collisions are possible.} and encodes a fraction of user's data bits. Additionally, the preamble identifies the unique features utilized in payload encoding such as scrambling and permutation sequences. These features facilitate the separation of users' payloads at the receiver side via multiple user detection (MUD). In the next sections we detail the operation of each system unit of the feed-forward link.

Consider a pool of $K_{\rm{tot}}$ users where a subset of $K_{\rm{a}}$ users are active at any given time slot. The $k^{th}$ active user encodes its $n$-bit data message using forward error-correction (FEC) encoder and obtains the codeword $\boldsymbol{v}^k = [\boldsymbol{v}_{\rm{p}}^k, \boldsymbol{v}_{\rm{d}}^k]$ of length $B$, $k=1,2,\cdots, K\stxt{a}$. In order to assist the receiver with the ability to decide on users' decoding success, cyclic redundancy check (CRC) is included into the  codeword. The vector $\boldsymbol{v}_{\rm{p}}^k$ is further encoded to form the preamble, while the vector $\boldsymbol{v}_{\rm{d}}^k$ is modulated to form the payload.

The vector $\boldsymbol{v}^k_{\rm{p}}$ that contains $B_{\rm{p}}$ bits is mapped into the preamble sequence $\boldsymbol{x}_{\rm{p}}^k \in \mathbb{C}^{N_{\rm{p}} \times 1}$ using a CS encoder. The sequence $\boldsymbol{x}\stxt{p}^k$ corresponds to a column in the sensing matrix $\mathbf{A} = [\boldsymbol{a}_1,\boldsymbol{a}_2,\cdots,\boldsymbol{a}_N],$ where the column index $\nu(k) = (\boldsymbol{v}^k_{\rm{p}})_{10}$ and $(.)_{10}$ indicates binary to decimal conversion,  $\boldsymbol{x}_{\rm{p}}^k = \boldsymbol{a}_{\nu(k)}$. 
The $ N_{\rm{p}} \times N$ matrix $\mathbf{A}$ where $N=2^{B_{\rm{p}}}$, composed of i.i.d. complex Gaussian entries, is chosen a priori and contains the pool of unitary preambles. These preamble sequences are selected by the active users according to the above-mentioned mapping. We will denote the set of active user indices by $\mathcal{K}_{\rm{a}} = \{\nu(k)\}, k=1,2,\cdots,K_{\rm{a}}$. In case several users select the same preamble sequence a collision occurs. Hence, the set $\mathcal{K}_{\rm{a}}$ may contain duplicate indices.

The payload is formed as follows. The remaining $B_{\rm{d}}$-bit sequence $\boldsymbol{v}^k_{\rm{d}}$ (where $B_{\rm{d}}+B_{\rm{p}}=B$) is encoded using repetition, permutation, and scrambling. Based on the index $\nu(k)$ the $k^{th}$ user selects a pair of payload scrambling and permutation sequences $(\boldsymbol{\pi}_{\nu(k)}, \boldsymbol{s}_{\nu(k)})$ from a pool of $2^{B_{\rm{p}}}$ scrambling and permutation sequence pairs used for payload encoding. Each bit of the data vector $\boldsymbol{v}_{\rm{d}}^k$ is repeated $M$ times and permuted using $\boldsymbol{\pi}_{\nu(k)}$. The resulting permuted bit replicas are used to form quadrature phase shift keying (QPSK) symbols which are then scrambled with $\boldsymbol{s}_{\nu(k)}$ (which is a sequence with random unitary complex entries) to obtain  $\boldsymbol{x}_{\rm{d}}^{k}$ that consumes $N_{\rm{d}}=N_{\rm{t}}-N_{\rm{p}}$ channel uses. 

\subsection{Preamble Decoder}
\label{sec:Receiver}
The receiver operates in two stages. First the CS decoder decodes the preambles and estimates, simultaneously, the set of the active user indices $\mathcal{K}\stxt{a}$ and channel coefficients $h_k$, $k \in \mathcal{K}\stxt{a}$ for each user. Note that once the activity detection (AD) estimates the preambles and resolves $\boldsymbol{v}^k\stxt{p}$ it also identifies the scrambling$-$permutation pairs $(\boldsymbol{\pi}_{\nu(k)}, \boldsymbol{s}_{\nu(k)})$ used for payload encoding. Then, the results of the preamble decoding are passed to the multi user detector to decode the payloads of the packets via an iterative parallel interference cancellation and data decoding algorithm.

The composite received signal for the preambles of the users' packets is given by
\begin{eqnarray}
\boldsymbol{y}_{\rm{p}} &=& \nonumber \sum_{k = 1}^{K\stxt{a}} h_k \boldsymbol{x}^k_{\rm{p}} + \boldsymbol{z}_{\rm{p}}\\
 &=& \mathbf{A} \boldsymbol{\mathfrak{h}} + \boldsymbol{z}_{\rm{p}}.
\end{eqnarray}
where $\boldsymbol{\mathfrak{h}}$ is a sparse ($2^{B\stxt{p}} \times 1$) vector of user activity and channel coefficients that has non-zero entries at $k \in \mathcal{K}\stxt{a}$.
The received signal $\boldsymbol{y}_{\rm{p}}$ contains preamble sequences $\boldsymbol{a}_k$ of the active users, weighted by the channel coefficients $h_k$. We consider user transmission over independent block Rayleigh channels with coefficients $h_k$. The additive white Gaussian noise (AWGN) at the receiver is denoted by $\boldsymbol{z}_{\rm{p}} \sim \mathcal{C}(0,N_0\mathbf{I}_{N\stxt{p}})$. 

The preamble CS decoder utilizes an iterative algorithm proposed in~\cite{tcom2021fading}  based on approximate message-passing (AMP)~\cite{donoho2009message}. Iteration $l$,  $l=0,1,2,\cdots$, starts with calculation of the residual noise 
\begin{eqnarray}
\boldsymbol{z}_{\rm{p}} ^{l} = \boldsymbol{y}_{\rm{p}}  - \mathbf{A}\widehat{\boldsymbol{\mathfrak{h}}}^{l-1} + \mathcal{O}(\boldsymbol{z}_{\rm{p}} ^{l}),
\end{eqnarray}
where $\widehat{\boldsymbol{\mathfrak{h}}}^{l}$ is the vector of all channel coefficient estimates at iteration $l$, $\widehat{\boldsymbol{\mathfrak{h}}}^{0}=\boldsymbol{0}$. The residual noise and interference vector $\boldsymbol{z}_{\rm{p}}^{l}$ is matched-filtered with the sensing matrix $\mathbf{A}$ and accumulated with the previous estimate, to find the initial estimate $\boldsymbol{r}^{l}$ of $\boldsymbol{\mathfrak{h}}$, that is 
\begin{eqnarray}
 \boldsymbol{r}^{l} &=& \boldsymbol{\widehat{\mathfrak{h}}}^{l} + \mathbf{A}^{\rm{*}}\boldsymbol{z}_{\rm{p}} ^{l}.
\end{eqnarray}
The estimate $\boldsymbol{r}^{l}$ is further improved to produce
\begin{eqnarray}
\widetilde{\boldsymbol{\mathfrak{h}}}_l = \eta(\boldsymbol{r}^{l},\tau_l^2),
\end{eqnarray}
using a soft denoiser $\eta(\cdot)$ (see definition in~\cite{FenglerC19})
where $\tau_l^2$ is the variance of the components of the vector $\boldsymbol{z}_{\rm{p}}$ computed via $\frac{||\boldsymbol{z}_{\rm{p}}||^2}{N_{\rm{p}}}$. 
Finally, thresholding with $c \tau_l$ is applied 
\begin{eqnarray}
\label{eqn:thresholding}
 \widehat{\mathfrak{h}}^{l+1}_j =
\begin{cases}
    \tilde{\mathfrak{h}}_j,&  |\tilde{\mathfrak{h}}_j|
    >  c \tau_l\\
    0,              & \text{otherwise}
\end{cases}, \quad j = 1,\cdots,2^{B\stxt{p}}
\end{eqnarray}
to reduce the number of false alarms and assist in convergence. 
The threshold multiplier $c$ is usually selected in the range $ 2 \leq c \leq 4$. This step is similar to the thresholding techniques that appear in many versions of AMP \cite{donoho2009message,maleki2013asymptotic}. After the final iteration, the receiver obtains the estimated indices of the active user preambles $\nu(k) \in \widehat{\mathcal{K}}_{\rm{a}}$ and their corresponding channel estimates $\widehat{h}_k = \tilde{\mathfrak{h}}_{\nu(k)}$, $k=1,2,\cdots,\widehat{K}\stxt{a}$, where $\widehat{K}\stxt{a}$ is the number of the detected users. The user activity detection errors consist of miss-detected users (where $\kappa \in \mathcal{K}_{\rm{a}}$ but $\kappa \not\in \widehat{\mathcal{K}}_{\rm{a}}$) or false alarms ($\kappa \not\in \mathcal{K}_{\rm{a}}$ but $\kappa\in \widehat{\mathcal{K}}_{\rm{a}}$). 

\subsection{Payload MUD}
\begin{figure}
\centering
\includegraphics[scale=1.3]{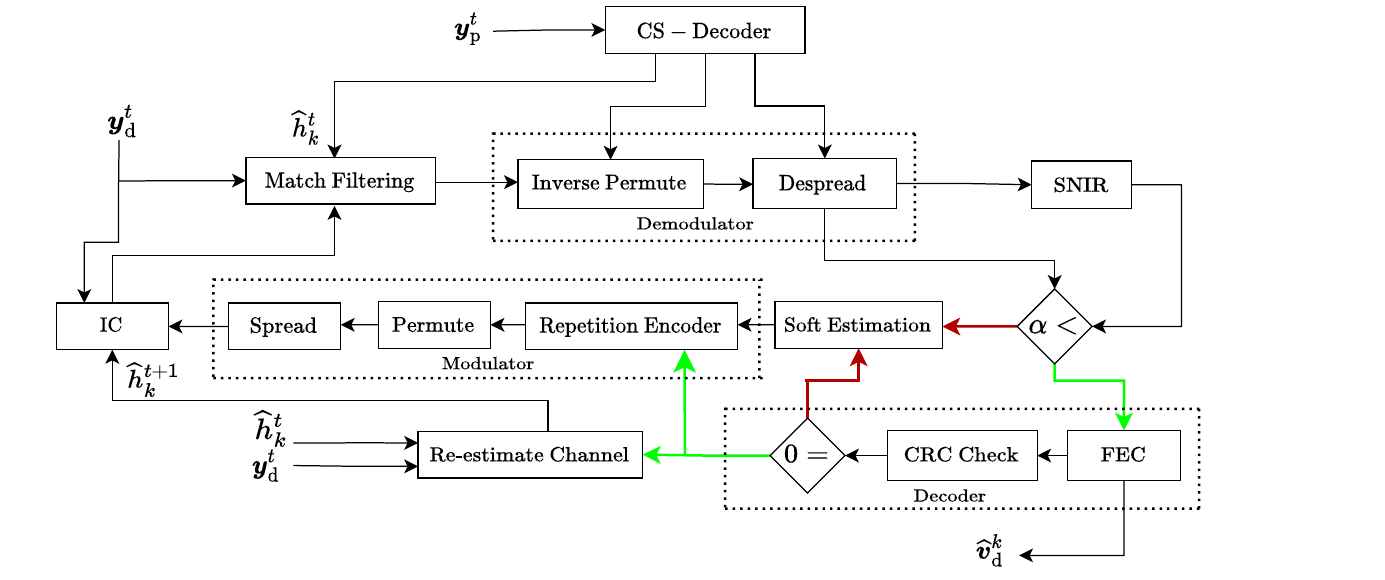}
\caption{Architecture of the payload MUD.}
\label{Fig:Rec}
\end{figure}
The MUD structure which includes both the interference cancellation and the FEC decoding, shown in Fig.~\ref{Fig:Rec} (see \cite{nassaji2021unsourced}), outputs a list $\mathcal{L}$ of the decoded codewords $\widehat{\boldsymbol{v}}^k$ for all detected users' signals.  
The composite received signal for the payloads is given by
\begin{eqnarray}
\boldsymbol{y}_{\rm{d}} = \sum_{k = 1}^{K\stxt{a}} {h}_{k}\boldsymbol{x}_{\rm{d}}^k +\boldsymbol{w}_{\rm{d}},
\end{eqnarray}
where $\boldsymbol{w}_{\rm{d}}$ is the iid Gaussian measurement noise.
At the beginning of the iterative MUD process,
the channel estimate $\hat{h}_k$ of each detected user $k$  
is utilized to produce the matched filtered received signal
\begin{eqnarray}
\boldsymbol{r}_k &=& \hat{h}_k^*\boldsymbol{y}\stxt{d} \nonumber \\
&=&|h_k|^2\boldsymbol{x}\stxt{d}^k+\sum_{\substack{{k}^\prime=1\\{k}^\prime\neq k}}^{K_{\rm{a}}}{\hat{h}^*_k h_{k'} \boldsymbol{x}_{\rm{d}}^{k'} }+\hat{h}^*_k\boldsymbol{w}_{\rm{d}}+e_k^* h_k\boldsymbol{x}\stxt{d}^k,
\label{eq:Cs5}
\end{eqnarray}
where $e_k = \hat{h}_k-h_k$. The MUD algorithm aims to iteratively reduce the inter-payload interference, and ameliorate user's payload signals $|h_k|^2 \boldsymbol{x}_{\rm{d}}^k$. 
Let us denote the signal of user $k$ at iteration $t-1$, $t=1,2,\cdots$, after the interference cancellation, by $\widetilde{\boldsymbol{r}}_k^{t-1}$, where $\widetilde{\boldsymbol{r}}_k^{0}=\boldsymbol{r}_k$. Based on $\widetilde{\boldsymbol{r}}_k^{t-1}$ the log-likelihood ratios (LLRs) of the transmitted data symbols at iteration $t$ are estimated via
\begin{eqnarray}
\widehat{\boldsymbol{\lambda}}_k^{t} = \frac{2 \boldsymbol{r}_k^{t-1}}{{\sigma}^2_{k,t-1}}.
\end{eqnarray}
where ${\sigma}^2_{k,t-1}$ is the power of the multi-user interference and noise observed by user $k$ at iteration $t-1$. The signal $\widehat{\boldsymbol{\lambda}}_k^{t}$ is multiplied by the scrambling sequence $\boldsymbol{s}_{\nu(k)}$, demodulated, and inverse-permuted using $\boldsymbol{\pi}^{-1}_{\nu(k)}$ to produce the LLRs of the payload data bits (see Fig.~\ref{Fig:Rec})
\begin{eqnarray}
\vartheta_{k,j}^{t} =\sum_{m=1}^{M}\widetilde{{\lambda}}_{k,j,m}^{t},
\end{eqnarray}
where $\widetilde{{\lambda}}^{t}_{k,j,m}$ is the LLR for the $m^{th}$ replica of the coded bit $v_{k,j,m}$, $j \in \{1,\hdots, B\stxt{d}\}$ of user $k$. Based on the LLRs we obtain soft bit estimates
\begin{eqnarray}
\label{bitEstimates}
\widetilde{v}_{k,j,m}^{t} =\text{tanh}\bigg(\sum_{\substack{m^\prime = 1\\ m^\prime \neq m}}^M \widetilde{{\lambda}}_{k,j,m^\prime}^{t}\bigg).
\end{eqnarray}
The bit estimates in (\ref{bitEstimates}) are used to re-modulate the user's signal with application of permutation, signature and a channel estimate to produce
\begin{eqnarray}
\boldsymbol{y}_{\rm{d}}^{k,t} = \widehat{h}_k^t \widehat{\boldsymbol{x}}_{\rm{d}}^{k,t}, 
\end{eqnarray}
which is utilized in interference cancellation.

When the signal-to-noise and interference ratio (SINR) at iteration $t$ is above a certain pre-set threshold $\alpha$ the error-correction decoder is activated at the receiver instead of repetition-based soft-bit estimation (\ref{bitEstimates}). The LLRs ${\vartheta}_{k,j}^{t}$ are passed to the input of the error-correction decoder together with the detected preamble data bits $\hat{\boldsymbol{v}}\stxt{p}^k$. 
In this paper polar codes with cyclic redundancy check (CRC) bits are utilized for data encoding, while successive cancellation list decoding is used at the receiver. 
If the decoded user's data codeword checks the cyclic redundancy check $\text{CRC} = 0$ for two consecutive iterations, it is assumed that the user's data is decoded successfully. The resulting re-modulated payload will further participate in hard interference cancellation and it's channel estimate update. The threshold $\alpha$ utilized here ranges from $(-20, -11)$ dB depending on the number of active users $K_{\rm{a}}$.


The bits output by the error-correction decoder (or soft bits (\ref{bitEstimates})) are utilized to form hard QPSK symbols, $\mathfrak{q}_{k,j}^t$ that are then repeated $M$ times, permuted, and scrambled, to re-modulate the data of the respective packet prior to the interference cancellation.  In order to aid the remaining users, the channels of the successfully decoded users are re-estimated using the decoded data as pilots. The part of the received signal that corresponds to the users, successfully detected and decoded  at iteration $t$ (set  $S_{\rm{a}}[t]$), is given by
\begin{eqnarray}
    \boldsymbol{y}_{\rm{c}} &=& \sum_{k\in S_{\rm{a}}[t]} h_k \boldsymbol{x}_{\rm{d}}^k + \boldsymbol{w}_t \nonumber \\
    &=& \mathcal{X}_t \boldsymbol{h}_t + \boldsymbol{w}_t,
\end{eqnarray}
where $\mathcal{X}_t$ is the matrix of the stacked successfully decoded data vectors acting as a pilot signal, and $\boldsymbol{w}_{t}$ is the signal of noise and interference from the other users (for which the interference cancellation is applied), with variance $\sigma_{\rm{NIP}}^2(t)$ (the noise and interference power (NIP)). The linear minimum mean square estimator (LMMSE) of the respective channel coefficients  $\boldsymbol{h}_t$ is given by
\begin{eqnarray}
    \widehat{\boldsymbol{h}}_t &=&  \mathcal{X}_t^* \bigg(\mathcal{X}_t\mathcal{X}_t^* + \sigma_{\text{NIP}}^2(t) \mathbf{I}  \bigg)^{-1} \boldsymbol{y}_{\rm{d}} \nonumber \\
    &=& \frac{\mathcal{X}_t}{\sigma_{\text{NIP}}^2(t)}\bigg( \mathbf{I} - \mathcal{X}_t\bigg(\mathbf{I} + \frac{\mathcal{X}_t^* \mathcal{X}_t}{\sigma_{\text{NIP}}^2(t)}\bigg)^{-1} \frac{\mathcal{X}_t^*}{\sigma_{\text{NIP}}^2(t)} \bigg) \boldsymbol{y}_{\rm{c}}. \nonumber 
\end{eqnarray}
The usefulness of the new channel estimates $\widehat{\boldsymbol{h}}_t$ is evaluated by the reduction of the noise and interference power (NIP) before and after the estimation. The NIP after the channel estimation is given by 
\begin{eqnarray}
\sigma_{\text{NIP}}^2(t^+) = ||\boldsymbol{y}_{\rm{c}} - \mathcal{X}_t\widehat{\boldsymbol{h}}_t||^2,
\end{eqnarray}
where by $t^+$ we denote an intermediate channel estimation step within the $t$th iteration. If $\sigma_{\text{NIP}}^2(t^+) < \sigma_{\text{NIP}}^2(t)$ then the new channel estimates reduce the overall NIP and are considered beneficial for the convergence process of the MUD. The new channel estimates can be combined along with the updated ones, but for simplicity we replace some channel estimates provided by the AD algorithm with the new updated ones. Finally, the interference cancellation is performed as follows 
\begin{equation}
\begin{aligned}
\tilde{\boldsymbol{r}}_k^t &= \hat{h}^*_{t,k}\boldsymbol{y}\stxt{d} - \sum_{\substack{{k}^\prime=1\\{k}^\prime\neq k}}^{\hat{K}\stxt{a}} \hat{h}^*_{t,k}\hat{h}_{t,k^\prime}\tilde{\boldsymbol{x}}\stxt{d}^{t,k'}  ,\ \\ 
\boldsymbol{\sigma}^2_{k,t}  &=\textrm{var} \left(\hat{h}^*_{t,k}\boldsymbol{y}\stxt{d} - \sum_{k'=1}^{\hat{K}_a} \hat{h}^*_{t,k}\hat{h}_{t,k^\prime}\tilde{\boldsymbol{x}}\stxt{d}^{t,k'}  \right)\ ,
\label{eq:Cs9}
\end{aligned} 
\end{equation}
in preparation for the next, $(t+1)$th iteration.
\subsection{Performance Metric}

By $\text{Pr}(E_k)$ we denote the probability of error in receiving user $k$'s message, where the error event $E_k = \{\boldsymbol{v}^k \not\in \mathcal{L}\} \cup \{\boldsymbol{v}^k = \boldsymbol{v}^i, k\not=i\} $ and $\mathcal{L}$ is the list of the decoded messages. 
Formally, the performance indicator PUPE is given by the average number of the incorrectly decoded user messages
\begin{eqnarray}
\label{def:PUPE}
\overline{P}_{\rm{e}} = \frac{1}{K_{\rm{a}}} \sum_{k = 1}^{K\stxt{a}} \text{Pr}(E_k).
\end{eqnarray}
This will include the error due to the missed users.

\section{Proposed Feedback mechanism}
\label{sec:FB_link}
\subsection{Baseline Feedback Systems}

Fundamentally, there are two basic approaches to perform the feedback in classical multi-user systems. In both approaches, the receiver explicitly informs a group of users about their detection/decoding status. In the positive-only feedback approach this group comprises of successful users (correctly detected and decoded users), while in the negative-only feedback this group consists of detected but incorrectly decoded users. In this section, we review these two basic feedback systems in the context of URA and use as a baseline to compare with our proposed feedback mechanism.  

\subsubsection{Positive-Only Feedback Approach}

In this approach, once the AD and MUD algorithms output the decoded users' data, the BS receiver constructs a feedback to acknowledge all successful users of their decoding status.\footnote{The decoding success is declared after checking the cyclic redundancy check (CRC) bits which are a part of the error-correction encoding.} For this purpose the receiver utilizes the preamble sequences $\boldsymbol{a}_k$ selected by the users on the feed-forward link to form the feedback signal
\begin{eqnarray}
\label{def:Pos-Only-FB}
\underline{\boldsymbol{x}} = \sum_{k \in S_{\rm{a}}} \boldsymbol{a}_k,
\end{eqnarray}
where the set of the successfully decoded active users is denoted by $S_{\rm{a}}$. We also define the set of the users that are detected but failed in decoding and denoted it by $F_{\rm{a}}$. The set of users that are missed (not detected) by the activity detection algorithm is denoted by $F_{\rm{md}}$.

Each active user processes the feedback signal $\underline{\boldsymbol{x}}$ broadcasted by the BS on the feedback link. If it detects its signature $\boldsymbol{a}_{k}$ in $\underline{\boldsymbol{x}}$ then a successful decoding acknowledgement is delivered to the $k$th user. Otherwise, the active user declares failure of detection/decoding and re-transmits its packet. Note that in the positive-only feedback approach the re-transmitting users are both the missed $and$ the incorrectly decoded users, i.e. both sets $F_{\rm{a}}$ and $F_{\rm{md}}$. Hence, all failed users are expected to be eventually recovered. Since most of the users are successful, high number of non-orthogonal preamble sequences is expected in (\ref{def:Pos-Only-FB}) that can cause high sequence interference and complicate the feedback detection process at the active users' side. 

\subsubsection{Negative-Only Feedback Approach}
In this approach, the feedback message contains only the preamble sequences of the failed users. The feedback signal in this case is given by
\begin{eqnarray}
\underline{\boldsymbol{x}} &=& \sum_{k \in F_{\rm{a}}} \boldsymbol{a}_k.
\end{eqnarray}

User $k$ processes the feedback message $\underline{\boldsymbol{x}}$ by looking for the presence of its preamble sequence $\boldsymbol{a}_{k}$ in the feedback message to decide whether it is intended by the BS for re-transmission. Unlike the positive-only approach, the feedback signal contains a smaller number of preamble sequences $\boldsymbol{a}_{k}$ but the complexity can still be significant in case of a large user pool. Note that the missed users (i.e. those in the set $F_{\rm{md}}$) will not re-transmit as they infer positive feedback acknowledgment. The latter causes higher feedback error compared to the positive-only feedback approach. In the next section  we propose a novel threshold-based feedback mechanism that simultaneously delivers positive $and$ negative feedback and, at the same time, addresses the majority of the successful users.  

\begin{figure}
    \centering
    \begin{tabular}{c c}
        \includegraphics{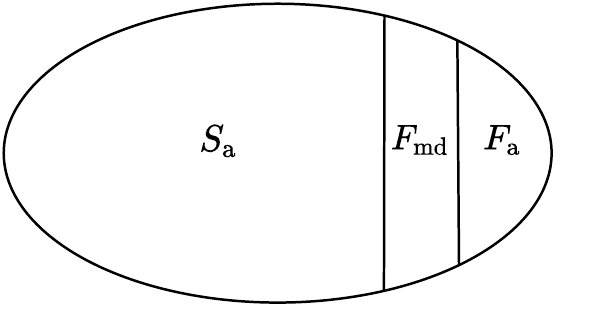}  &     \includegraphics{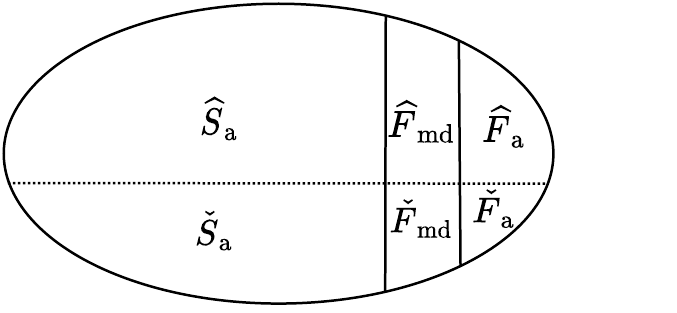} \\
        (a) & (b)
    \end{tabular}
    \caption{(a) Venn's Diagram of the user sets after feed-forward detection and decoding processes. (b) Further user classification based on user channels gains $h_k$ and the threshold $\widetilde{c}\tau$.}
    \label{fig:usersClassification}
\end{figure}

\subsection{Single-Threshold Feedback}

Fig.~\ref{fig:usersClassification}~(a) shows the three classes of the  active user messages after the AD and MUD processing on the feed-forward link. Some users are missed by the AD (set $F_{\rm{md}}$), some of the detected users are not decoded successfully by the MUD (set $F_{\rm{a}}$), and, finally, some users are both successfully detected and decoded (set $S_{\rm{a}}$). 

We recall that on the feed-forward link the AMP channel estimation algorithm uses the received signal and the preamble sequence dictionary matrix $\mathbf{A}$ to iteratively produce the estimates of the user channel gains $h_k$ (corresponding to the preamble sequences $\boldsymbol{a}_k$, $k \in \{1,2,\cdots,2^{B_{\rm{p}}}\}$ where $2^{B_{\rm{p}}}$ is the number of possible preambles). At each iteration, the AMP algorithm applies component-wise thresholding with $c\tau$ (as shown in Section~\ref{sec:Receiver} (\ref{eqn:thresholding})) to its estimated channels vector $|\widehat{\boldsymbol{h}}|$ in order to reject small entries which are perceived to be the noise rather than active users' channels. This reduces the presence of false alarms which can otherwise cause AMP to diverge. 
The threshold is applied actively to suppress noise spikes assuming that all detected entries with magnitude below $c \tau$ are caused by the noise. The majority of AD algorithms follow the same thresholding strategy, and, therefore, our feedback strategy is applicable to a general thresholding AD algorithm. 

We assume that the channel between a user and the BS satisfies the reciprocity condition, i.e. the channel gains on the feed-forwrd and feedback links are the same\footnote{In fact the proposed technique can work with a milder assumption that just the absolute value of the channel gain is constant (approximately).}. Then, an active user $k$ is able to learn whether it has been missed by the AD algorithm on the feed-forward link if it knows the threshold $c \tau$ of the AD algorithm and estimate of its channel $\hat{h}_k$ on the feedback link. That is, if $|\hat{h}_k| < c \tau$ then the $k$th user was most likely missed by the AD algorithm on the feed-forward link, and is required to re-transmit its message.  While users below the AD threshold $c \tau$ are definitely missed by the AD, some users above the threshold are also missed due to noise and estimation errors. Negative feedback needs to be provided to these users, as well. Note that instead of using the threshold $c\tau$ of the AD algorithm we can use a generalized threshold $\widetilde{c} \tau$, where $c < \widetilde{c}$, and $\widetilde{c}$ can be chosen such that all users above $\widetilde{c}$ are detected correctly by the AMP and $all$ users below it are missed. 

Fig.~\ref{fig:usersClassification}~(b) depicts active user categorization with respect to the threshold $\widetilde{c}\tau$. The sets $\widehat{F}_{\rm{md}}$ and $\widecheck{F}_{\rm{md}}$ correspond to the missed users with channels \textit{above} ($|\hat{h}_k| \geq  \tilde{c} \tau$) and \textit{below} ($|\hat{h}_k| <  \tilde{c} \tau$) the threshold respectively. The set $\widehat{F}_{\rm{a}}$ denotes the set of the failed users (detected by the AD but incorrectly decoded) above the threshold, while $\widecheck{F}_{\rm{a}}$ denotes the set of the failed users below the threshold. Finally, the set of successfully detected and decoded users above the threshold is denoted by $\widehat{S}_{\rm{a}}$, while the set of successful users below the threshold is denoted by $\widecheck{S}_{\rm{a}}$. In the next sections we first show how the feedback is designed to instruct each user group with the necessary action to \textit{re-transmit} or \textit{not to re-transmit} and then compute the corresponding cost functions and expected reduction in PUPE. Additionally, we design categorization of the users around two thresholds that can further reduce computational demands on active user resources.  


Assuming that all active users already acquired the knowledge of the threshold $\widetilde{c}\tau$ and channels $h_k$, the user groups below the threshold $\widecheck{F} = \{\widecheck{F}_{\rm{md}}, \widecheck{F}_{\rm{a}} \}$ can re-transmit without reception of negative feedback messages, addressed specifically to them. The users in group $\widecheck{F}$ would infer that they have been missed by the AD algorithm and re-transmit. We call this an \textit{implicit} negative feedback. The problem with such approach, however, is that the users in the set $\widecheck{S}_{\rm{a}}$ will re-transmit as well, since they also infer negative feedback. Therefore, direct positive feedback messages are provided to the users in $\widecheck{S}_{\rm{a}}$ so that they don't re-transmit. Once the users above the threshold learn their channels and the threshold they immediately infer positive feedback. The failed users in the set $\widehat{F}_{\rm{a}}$ will receive an explicit negative feedback commanding them to re-transmit. The PUPE of the system before the feedback equals
\begin{eqnarray}
P_{\rm{e}} = \frac{|\widecheck{F}| + |\widehat{F}|}{K_{\rm{a}}},
\end{eqnarray}
where $\widehat{F}= \{ \widehat{F}_{\rm{md}},\widehat{F}_{\rm{a}} \}$. In the case of optimal threshold selection and perfect user's channel knowledge, the error after feedback processing reduces to
\begin{eqnarray}
\underline{P}_{\rm{e}}^{\rm{s}} = \frac{ |\widehat{F}_{\rm{md}}|}{K_{\rm{a}}}.
\end{eqnarray}
This is because the user group $\widehat{F}_{\rm{md}}$ infers positive feedback and does not re-transmit despite the fact that these users fail on the feed-forward link. Note that it is highly unlikely for a user to fall into this group, especially if the threshold $\tilde{c}\tau$ is increased compared to the AD threshold.

In order to allow active users to acquire both channel and threshold knowledge to process the feedback as explained above, the BS constructs the feedback packet that carries the feedback information according to 
\begin{eqnarray}
\label{singlThresholdPacket}
\boldsymbol{x}_{\rm{f}} &=&  \begin{bmatrix} \boldsymbol{p} & \widetilde{c}\tau \boldsymbol{p} & \underline{\boldsymbol{x}}
\end{bmatrix}^T.
\end{eqnarray}
This packet is broadcasted to all users in the feedback slot. The first part of the feedback packet is a short pilot signal $\boldsymbol{p}$ utilized by the active users to obtain estimates $\widehat{h}_k$ of their channels. The second part is the broadcast weighted pilot signal $\boldsymbol{p}$ scaled by the threshold $\widetilde{c}\tau$. This will allow the users to estimate the threshold $\widetilde{c}\tau$ (Alternatively, a quantized and encoded numerical value of $\widetilde{c}\tau$ can be included into the feedback packet). The third part, i.e. $\underline{\boldsymbol{x}}$, is a superposition of the preamble sequences $\boldsymbol{a}_k$ of the users for which an explicit positive or negative feedback is provided. The failed users in the set $\widehat{F}_{\rm{a}}$  will get a negative feedback $ - \boldsymbol{a}_k$ commanding them  to re-transmit, while the users which are below the threshold $\widetilde{c} \tau$ but are still detected and decoded correctly, i.e., $\widecheck{S}_{\text{a}}$, will get a positive feedback $ + \boldsymbol{a}_k$ to avoid re-transmission. Formally we can construct\footnote{Note that $-1$ or $+1$ in front of the targeted feedback is optional since we already know that the users in $\widehat{F}_{\rm{a}}$ look for a possible negative feedback message and users in $\widecheck{S}_{\rm{a}}$ for a possible positive feedback message.},
\begin{eqnarray}
\underline{\boldsymbol{x}} &=&  \sum_{k \in \widehat{F}_{\rm{a}} \cup \widecheck{S}_{\rm{a}}}  \boldsymbol{a}_{k}.
\end{eqnarray}

The $k^{th}$ active user receives the feedback signal 
\begin{eqnarray}
\boldsymbol{y}_k = h_k \boldsymbol{x}_{\rm{f}} + \boldsymbol{z}_k,
\end{eqnarray}
where  $\boldsymbol{z}_k$ is the vector of AWGN, and starts  processing the first part (see (\ref{singlThresholdPacket})), i.e., the received pilot signal 
\begin{eqnarray}
\boldsymbol{y}_k^{(1)} = h_k \boldsymbol{p} + \boldsymbol{z}_k^{(1)} ,
\end{eqnarray}
to estimate its channel. Algorithm~\ref{Alg1} describes the decision process at the user end. The linear minimum mean square estimator (LMMSE) of the user's channel is given by 
\begin{eqnarray}
\label{LMMSE}
\widehat{\underline{h}}_k &=& \boldsymbol{p}^*(\boldsymbol{p}\boldsymbol{p}^* + \sigma_{z}^2 \mathbf{I})^{-1} \boldsymbol{y}_k^{(1)} 
= \boldsymbol{p}^* \big( \mathbf{I} - \frac{ \boldsymbol{p}\boldsymbol{p}^*}{\sigma_z^2 +  |\boldsymbol{p}|^2} \big)\frac{\boldsymbol{y}_k^{(1)}}{\sigma_{z}^2}, 
\label{eq:minv}
\end{eqnarray}
where the matrix inversion lemma is utilized to simplify the calculations and avoid a matrix inversion. In (\ref{eq:minv}), $\sigma_{z}^2$ is the noise power, which can be estimated using silence periods. The $k^{th}$ user's channel estimate obtained by the user itself on the feedback link is denoted by $\widehat{\underline{h}}_k$, while the estimate of the same channel (due to reciprocity assumption) performed by the BS is denoted by $\widehat{h}_k$. Note that the pilot-dependent part of the feedback signal (such as $\boldsymbol{p}\boldsymbol{p}^*$ and $|\boldsymbol{p}|^2$) can be pre-calculated and saved at the user devices to further reduce the complexity. Following the channel estimation, the users recover the threshold from the received signal 
\begin{eqnarray}
\boldsymbol{y}_k^{(2)} &=& h_k \widetilde{c} \tau \boldsymbol{p} + \boldsymbol{z}_k^{(2)},
\end{eqnarray}
using the LMMSE 
\begin{eqnarray}
\widehat{\tau \tilde{c}} &=& \widehat{\underline{h}}_k \boldsymbol{p}^* \big( \mathbf{I} - \frac{|\widehat{\underline{h}}_k|^2 \boldsymbol{p}\boldsymbol{p}^*}{\sigma_z^2 + |\widehat{\underline{h}}_k|^2 |\boldsymbol{p}|^2} \big)\frac{\boldsymbol{y}_k^{(2)}}{\sigma_{z}^2}.
\end{eqnarray}
In order to finally decide on re-transmission, each user processes the targeted acknowledgement part of the packet $\underline{\boldsymbol{x}}$, i.e., the third part of the received signal 
\begin{eqnarray}
\boldsymbol{y}_k^{(3)} &=& h_k \underline{\boldsymbol{x}} + \boldsymbol{z}_k^{(3)}  = h_k \boldsymbol{a}_k + h_{k} \sum_{\acute{k} \neq k}  \boldsymbol{a}_{\acute{k}} + \boldsymbol{z}_k^{(3)}.
\end{eqnarray}
\begin{algorithm}[t]
 \textbf{Input: }{User's channel estimate $\underline{\widehat{h}}_k$,
 feedback decision threshold $\widetilde{c}\tau,$
 correlation with user's preamble $\gamma_k,$ correlation comparison threshold $\bar{\gamma}$}. \\
 \textbf{Output: }{re-transmit \textit{or} not to re-transmit decision}\\
 \eIf{$|\gamma_k| > \bar{\gamma}$ }{\eIf{$|\underline{\widehat{h}}_k| < \widetilde{c}\tau$}{do not re-transmit}{re-transmit}}
 {\eIf{$|\underline{\widehat{h}}_k| \geq \widetilde{c}\tau$}{do not  re-transmit}{re-transmit}}
 \caption{User's Decision Procedure For Single-Threshold Feedback}
 \label{Alg1}
\end{algorithm}
The $k^{th}$ user correlates $\boldsymbol{y}_k^{(3)}$ with its signature $\boldsymbol{a}_{k}$ to produce the statistics
\begin{eqnarray}
\label{gamma_k}
\gamma_k &=& \frac{\boldsymbol{a}_k^* \boldsymbol{y}_k^{(3)}}{\widehat{\underline{h}}_k}.
\end{eqnarray}
If $|\gamma_k| > \bar{\gamma}$, where $\bar{\gamma}$ is a certain preset threshold, the user decides that it detects a presence of the positive or negative targeted feedback. The user may also use $\text{sign}(\text{real}(\gamma_k))$ to infer the feedback type. The described feedback process incurs a complexity cost for both BS and active users. The BS cost is defined as the number of users with preamble sequences included in $\underline{\boldsymbol{x}}$ and is equal to 
\begin{eqnarray}
\mathcal{C}_{\rm{BS}}^{\rm{s}} &=& |\widecheck{S}_{\rm{a}}| + |\widehat{F}_{\rm{a}}|.
\end{eqnarray}
Note that $\mathcal{C}_{\rm{BS}}^{\rm{s}}$ is threshold-dependant. The lower the cost, the easier it is for the users to detect the intended targeted feedback because of the smaller amount of the multi-user interference in the signal $\boldsymbol{y}_k^{(3)}$.
The processing cost at the active user side is defined as the number of users that have to process the {\em entire} feedback packet $\boldsymbol{x}_{\rm{f}}$. For the case of single-threshold feedback approach it is given by 
\begin{eqnarray}
\mathcal{C}_{\rm{UE}}^{\rm{s}} &=& K_{\rm{a}}, 
\end{eqnarray}
and is threshold-independent, since all active users have to process the entire signal $\boldsymbol{x}_{\rm{f}}$. 
In the next section, we propose an improved double-threshold feedback approach that also reduces the users' processing cost. 

\subsection{Double-Threshold Feedback}
\begin{figure}[t]
    \centering
    \includegraphics{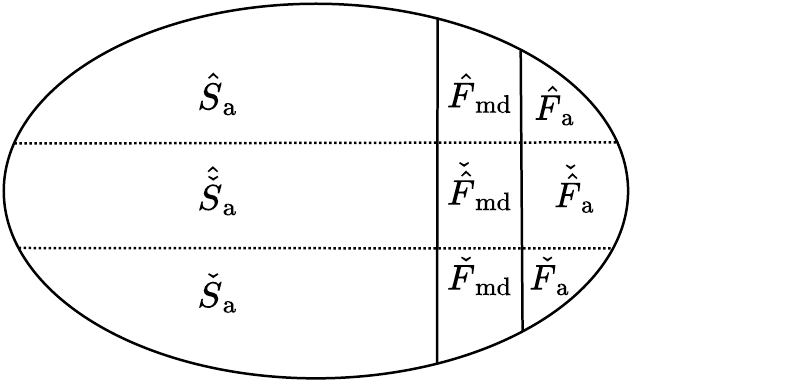}
    \caption{User classification after the AD and  MUD for the case of double-threshold feedback with thresholds $\tilde{c}_1 \tau$ and $\tilde{c}_2 \tau$. Users with channel amplitude above $\tilde{c}_1 \tau$ infer positive feedback, while the users with channel amplitude below $\tilde{c}_2 \tau$ will negative feedback.}
    \label{fig:2thresholdSets1}
\end{figure}
Using the single-threshold feedback mechanism all the active users process the entire received feedback signal ($[\boldsymbol{y}_k^{(1)}, \boldsymbol{y}_k^{(2)}, \boldsymbol{y}_k^{(3)} ]$) to decide whether to re-transmit or not. Since most of the users are detected and decoded correctly on the feed-forward link ($\widehat{S}_{\rm{a}}$), it is more efficient to acquire their status after processing only a small portion of the feedback signal. This reduces the cost of the feedback processing significantly at a small additional signalling overhead. To save the power at the users' side, we propose a double-threshold feedback mechanism that takes into account the probability of high detection/decoding success of a user packet. 

Fig.~\ref{fig:2thresholdSets1} shows user categorization imposed by two thresholds $\widetilde{c}_1 \tau$ and $\widetilde{c}_2 \tau$, where $\widetilde{c}_2 > \widetilde{c}_1$. Unlike the single-threshold case, all users with $|h_k| < \widetilde{c}_1 \tau$, i.e., in groups $\widecheck{F}_{\rm{md}}$, $\widecheck{F}_{\rm{a}}$ and $\widecheck{S}_{\rm{a}}$ always re-transmit without checking on targeted feedback messages in $\underline{\boldsymbol{x}}$. Similarly, all users with $|h_k| > \widetilde{c}_2 \tau$, i.e., in groups $\widehat{F}_{\rm{md}}$, $\widehat{F}_{\rm{a}}$ and $\widehat{S}_{\rm{a}}$, refrain from re-transmission since they infer positive feedback. Only users between the two thresholds (i.e. with $\widetilde{c}_1 \tau < |h_k| < \widetilde{c}_2 \tau$)  will process the entire feedback signal including the targeted feedback part $\underline{\boldsymbol{x}}$. 
The feedback signal $\boldsymbol{x}_{\rm{f}}$ for the double-threshold approach has the format
\begin{eqnarray}
\boldsymbol{x}_{\rm{f}} &=&   \begin{bmatrix} \boldsymbol{p} & \widetilde{c}_2 \tau \boldsymbol{p} & \widetilde{c}_1 \tau \boldsymbol{p} & \underline{\boldsymbol{x}}
\end{bmatrix}^T,
\end{eqnarray}
\begin{algorithm}
 \textbf{Input: }{User's channel estimate $\underline{\widehat{h}}_k$,
 upper and lower feedback decision thresholds $\widetilde{c}_2\tau,\widetilde{c}_1\tau$,
 correlation with user's preamble $\gamma_k,$ correlation comparison threshold $\bar{\gamma}$.}\\
 \textbf{Output: }{re-transmit \textit{or} not to re-transmit decision.} \\
 \eIf{$|\underline{\widehat{h}}_k| > \widetilde{c}_2 \tau$}{do not re-transmit and stop further processing of the feedback signal}{process further to find $\widetilde{c}_1 \tau$ \\ 
 \eIf{$|\underline{\widehat{h}}_k| < \widetilde{c}_1 \tau$}{re-transmit and stop further processing of the feedback signal}{process further to find $\gamma_k$ \\ \eIf{$|\gamma_k| > \bar{\gamma}$}{do not re-transmit}{re-transmit}}}
 \caption{User's Decision Procedure For Double-Threshold Feedback}
 \label{Alg2}
\end{algorithm}
where signal $\underline{\boldsymbol{x}}$ is given by
\begin{eqnarray}
\underline{\boldsymbol{x}} = \sum_{k \in \widecheck{\widehat{S}}_{\text{a}}} \boldsymbol{a}_{k}. 
\end{eqnarray}
where the set of successful users between the two thresholds is denoted by $\widecheck{\widehat{S}}_{\rm{a}}$. Signal $\underline{\boldsymbol{x}}$ instructs the users in $\widecheck{\widehat{S}}_{\rm{a}}$ not to re-transmit. The sets of missed $\widecheck{\widehat{F}}_{\rm{md}}$ and  failed users $\widecheck{\widehat{F}}_{\rm{a}}$ between the two thresholds infer implicit negative feedback and re-transmit. As in the case for the single-threshold feedback, all users utilize the first part of the received feedback signal (pilot) to estimate their channels. All users also process the second part of the feedback signal to acquire $\widetilde{c}_2 \tau$. The thresholds are selected in such a way that after this stage most of the users infer positive feedback and stop here. Only users that fall below the second threshold start processing the third block. After acquiring $\tilde{c}_1\tau$, even fewer users (in sets $\widecheck{\widehat{S}}_{\rm{a}}$, $\widecheck{\widehat{F}}_{\rm{md}}$, and $\widecheck{\widehat{F}}_{\rm{a}}$) engage in processing  of the final part $\underline{\boldsymbol{x}}$. 

The cost of the double-threshold feedback at the BS side is given by
\begin{eqnarray}
\mathcal{C}_{\rm{BS}}^{\rm{d}} &=& |\widecheck{\widehat{S}}_{\rm{a}}|,
\end{eqnarray}
and is threshold-dependent. The user's cost however, drops to 
\begin{eqnarray}
\mathcal{C}_{\rm{UE}}^{\rm{d}} &=& |\widecheck{\widehat{S}}_{\rm{a}}| + |\widecheck{\widehat{F}}_{\rm{a}}| + |\widecheck{\widehat{F}}_{\rm{md}}|,
\end{eqnarray}
which is also threshold dependent and is significantly smaller than the user's cost for the single-threshold method which is the entire $K_{\rm{a}}$. In the single-threshold method, the feedback errors (in case of ideal feedback reception) are caused by the set of missed users above the threshold, i.e., by the set $\widehat{F}_{\rm{md}}$ only. In the double-threshold method the detected/undecoded users above $\tilde{c}_2 \tau$ add to the feedback error, that becomes
\begin{eqnarray}
\underline{P}_{\rm{e}}^{\rm{d}} &=& \frac{|\widehat{F}_{\rm{md}}| + |\widehat{F}_{\rm{a}}|}{K_{\rm{a}}} \\
&=& \underline{P}_{\rm{e}}^{\rm{s}} + \frac{|\widehat{F}_{\rm{a}}|}{K_{\rm{a}}}. \nonumber 
\end{eqnarray}
We note that the overall PUPE of the system with double-threshold feedback is slightly higher than that of the single-threshold approach for the same parameter settings. Table~\ref{tab:my_label} summarizes and compares the costs the feedback approaches.

\subsection{Feedback with Partial Preamble Sequences}

In order to conserve the available channel resources, only a portion of each user's preamble sequence can be used to construct the feedback signal $\underline{\boldsymbol{x}}$, instead of the full preamble sequence. This shortens the feedback packet (saving channel resources) and further reduces the energy spent at the user's side to process the feedback. In the next section we compare the performance of the proposed feedback system, and the baseline systems for both full and partial sequence settings.  

\begin{table}
    \centering
    \begin{tabular}{|c|c|c|}
    \hline
     Feedback approach& $\mathcal{C}_{\rm{BS}}$ 
         & $\mathcal{C}_{\rm{UE}}$ \\
         \hline 
         Single-threshold & $|\widecheck{S}_{\rm{a}}| + |\widehat{F}_{\rm{a}}|$ & $K_{\rm{a}}$ \\
         \hline 
         Double-threshold & $|\widecheck{\widehat{S}}_{\rm{a}}| + |\widecheck{\widehat{F}}_{\rm{a}}|$ & $|\widecheck{\widehat{S}}_{\rm{a}}| + |\widecheck{\widehat{F}}_{\rm{a}}| + |\widecheck{\widehat{F}}_{\rm{md}}|$ \\
         \hline 
    \end{tabular}
    \caption{Cost of the feedback (in terms of user sets) at BS and UE side.}
    \label{tab:my_label}
\end{table}


\section{Numerical Results}
\label{sec:NumericalResults}
In this section we study the effects of presented feedback algorithms on improving the reliability and reducing the required operational $E_{\rm{b}}/N_{\rm{o}}$ of a URA system. We show that the threshold-based feedback can significantly enhance the reliability (reducing PUPE), without providing an explicit feedback to all URA users. We demonstrate that the single-threshold approach has the complexity of the negative-only feedback approach but with the gain of the positive-only feedback. More importantly, the proposed feedback conserves the spectral efficiency of the system. Additionally, the proposed feedback mechanism allows a system designer to reduce the overall $E_{\rm{b}}/N_{\rm{o}}$ required to attain the required overall reliability (e. g. $\rm{PUPE} = 0.05$).

\subsection{Feedback error and cost with perfect feedback reception}
First we focus on the feedback performance in terms of cost and the number of users that receive erroneous feedback. We consider a feed-forward link defined in Section~\ref{se:FFLink} followed by a perfect feedback where we assume that each user estimates its channel, the threshold $\widetilde{c}\tau$, and recovers its preamble from $\underline{\boldsymbol{x}}$ (if it is present) without any error. The total number of feed-forward channel uses equals $N = 7500$, the preamble length equals $N_{\rm{p}} = 2000$. Each user transmits $n = 100$ information bits encoded using the shortened Hamming code $(109,100)$, shortened from the double-extended Hamming code used in \cite{400ZR}. $B_{\rm{p}} = 15$ bits are encoded into the preamble and the remaining $B_{\rm{d}}$ bits into the payload. We assume that after the detection and decoding the receiver acquires the information about the correctness of the decoding from a higher layer. In Section~\ref{seq:NumB} we drop this assumption and utilize Polar coded FEC with included CRC. 
In Section~\ref{seq:NumB} we drop the assumption of the perfect feedback reception as well, and employ the fully integrated feed-forward/feed-back system to confirm our findings and present additional results. Block fading Rayleigh channel model is considered with the channel reciprocity assumption where  channels of the feed-forward and feedback link are the same for each user.

\begin{figure}
    \centering
    \includegraphics[scale=0.8]{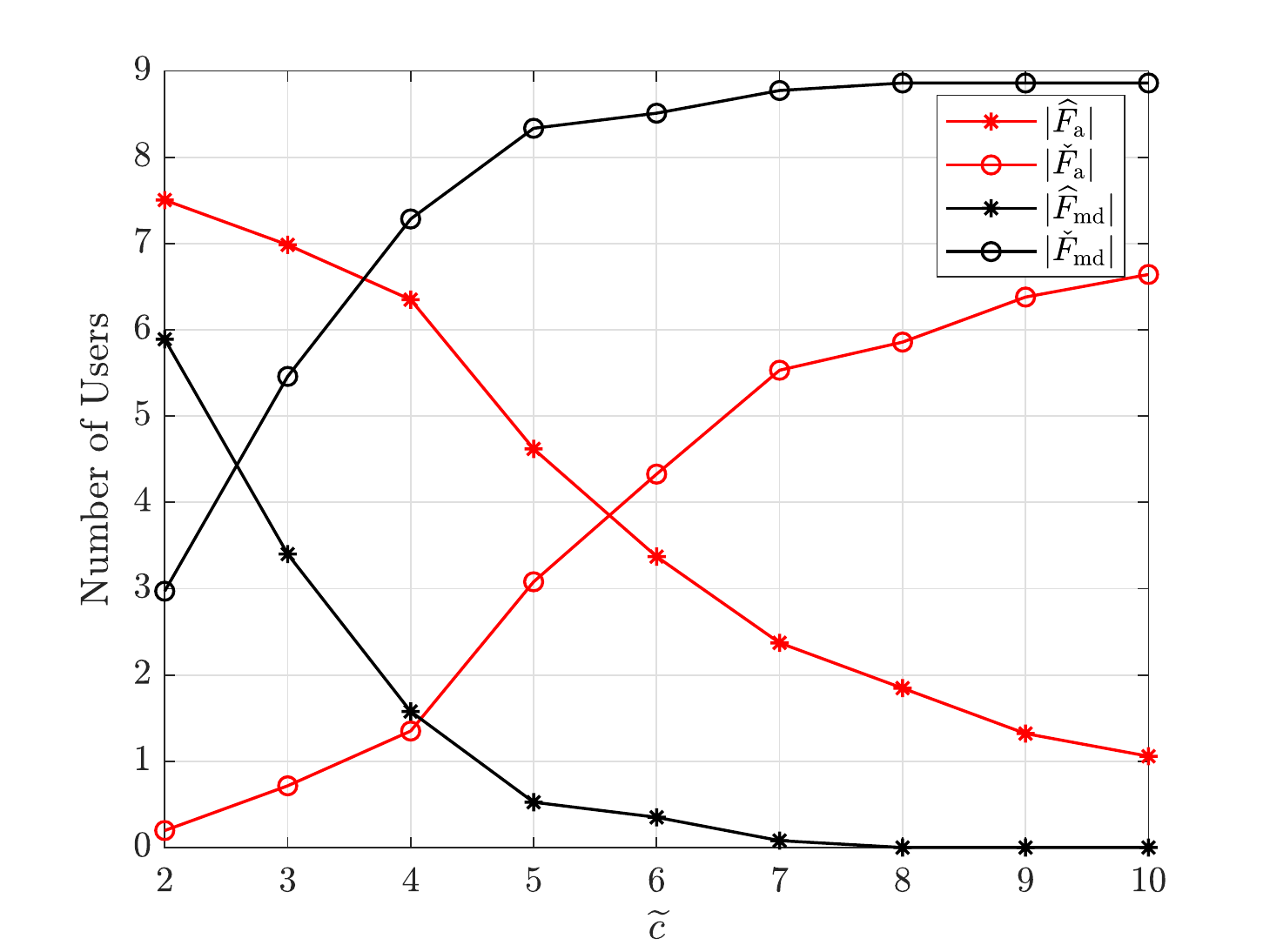}
    \caption{The numbers of users in the sets targeted by the single-threshold feedback for the Hamming-coded URA system with perfect feedback.}
    \label{fig:thresholdFeedback300UsersResults}
\end{figure}

Fig.~\ref{fig:thresholdFeedback300UsersResults} shows the average number of users in the sets depicted in Fig.~\ref{fig:usersClassification} (b), as a function of the threshold $\widetilde{c}\tau$ for a system with $K_{\rm{a}} = 300$ active users and SNR $\frac{E_{\rm{b}}}{N_o} = 17.6$ dB of the feed-forward link.  
Since it is likely that a failed user has a low channel gain, the number of failed users below the threshold $|\widecheck{F}_{\rm{a}}|$ increases as the threshold increases. 
Conversely, the number of failed users above the threshold $|\widehat{F}_{\rm{a}}|$ decreases steadily. We observe the same trend with the missed user sets, where the users with low channel gains are also likely to be missed by the AD algorithm. The figure demonstrates that high threshold multiplier $\widetilde{c}$ can be used to eliminate the feedback error almost entirely, since the number of missed users above threshold $|F_{\rm{md}}|$ is the only source of the feedback error in this perfect feedback link scenario. 

Now we consider the positive-only feedback system with the same parameters for comparison. Unlike the single-threshold feedback which targets successful users implicitly, the positive-only feedback explicitly targets users in sets $\widecheck{S}_{\rm{a}} + \widehat{S}_{\rm{a}} = S_{\rm{a}}$. The failed $F_{\rm{a}}$ and missed $F_{\rm{md}}$ users receive implicit negative acknowledgement. 
The cost of the positive-only feedback (at the BS side), defined as the number of users targeted by the feedback, is given by
\begin{eqnarray}
\mathcal{C}_{\rm{BS}}^{+} = (1 -\overline{P}_{\rm{e}}) K_{\rm{a}} = |S_{\rm{a}}| = 282.8 \text{ users ,}
\label{eq:pos94}
\end{eqnarray}
where $\overline{P}_{\rm{e}}$ is the actual PUPE of the feed-forward link. 
The value of $\mathcal{C}_{\text{BS}}^{+}$ is considerably higher than the cost of the proposed single-threshold approach $\mathcal{C}_{\rm{BS}}^{\rm{s}}$, which is dominated by $|\widecheck{S}_{\rm{a}}| \ll |S_{\rm{a}}|$. For the system we consider, the threshold multiplier $\widetilde{c} \geq 8$ leads to the single-threshold feedback cost equal to
\begin{eqnarray}
\mathcal{C}_{\rm{BS}}^{\rm{s}} &=& |\widehat{F}_{\rm{a}}| + |\widecheck{S}_{\rm{a}}|
\approx 0.8 + 30 = 30.8 \text{ users },
\end{eqnarray}
which is approximately $10\%$ of the total active user population $K_{\rm{a}}$ (worst case cost) compared to $94\%$ of $K_{\rm{a}}$ in case of positive-only feedback (\ref{eq:pos94}).


Such substantial cost reduction ($\mathcal{C}_{\rm{BS}}^{\rm{s}} \ll \mathcal{C}_{\rm{BS}}^{+}$) enables design and utilization of simple feedback processing algorithms at the active user side. Simplicity of the feedback detection algorithms is usually a requirement since most user devices have limited resources and cannot utilize the same detection algorithms as these available at the BS. 


\begin{table}[]
    \centering
    \begin{tabular}{|c||c|c|c||c|c|c||c|c|c|}
    \hline 
         $K_{\rm{a}}$& \multicolumn{3}{c||}{$50$} & \multicolumn{3}{c||}{$150$}  
         & \multicolumn{3}{c|}{$300$} \\
         \hline 
         $\text{FB Type}$ & $\mathbf{+}$ & $\mathbf{-}$ & $\mathbf{\sim}$ & $\mathbf{+}$ & $\mathbf{-}$ & $\mathbf{\sim}$ & $\mathbf{+}$ & $\mathbf{-}$ & $\mathbf{\sim}$ \\
         \hline 
         $\underline{P}_{\rm{e}}$ & $0$ & $0.04$ & $0 - 0.003$ & $0$ & $0.05$ & $0 - 0.003$ & $0$ & $0.07$ & $0 - 0.003$ \\
         \hline 
         $\mathcal{C}_{\rm{BS}}$ & $45$ & $6.67$ & $1.903 - 0.125$ & $135$ & $10.75$ & $9.7 - 1.8$ & $270$ & $8$ & $21.53- 0$ \\
         \hline 
    \end{tabular}
    \caption{The BS cost and the feedback error of the proposed single-threshold feedback approach (denoted by $\mathbf{\sim}$) compared to the costs of the positive ($\mathbf{+}$) and negative ($\mathbf{-
    }$) feedback baseline approaches for the systems with $K_{\rm{a}}=50, 150, 300$ active users.}
    \label{fig:FBcomp}
\end{table} 
 
In the negative-only feedback approach, only failed users are notified about their status and the rest of the users will infer successful decoding including the users missed by the AD. Therefore, only failed users re-transmit while all missed users are not able to re-transmit and also contribute to the feedback error. 
The BS cost of the negative-only feedback is given by
\begin{eqnarray}
\mathcal{C}_{\text{BS}}^{-}  = |F_{\rm{a}}| = 8 \text{ users } ,
\end{eqnarray}
and the corresponding PUPE feedback error is given by
\begin{eqnarray}
\underline{P}_{\rm{e}}^{\rm{-}} &=&  \frac{|F_{\rm{md}}|}{K_{\rm{a}}} = 8.9/300 \approx 0.03.
\end{eqnarray}
By setting $\widetilde{c} = 2$, the cost of the proposed single-threshold approach reduces to $|\widehat{F}_{\rm{a}}| + |\widecheck{S}_{\rm{a}}| = 7.6$ users with $\underline{P}_{\rm{e}}^{\rm{s}} \approx 0.02$. 
Table~\ref{fig:FBcomp} presents a comparison between the BS cost and feedback PUPE of the baseline approaches and the proposed single-threshold approach for different numbers of users under the same parameter setting. We can see that the proposed feedback approach presents a very significant cost reduction at the expense of a very minor feedback error compared to the positive-only feedback.  

Note that since each user has to process the entire feedback signal in all these approaches the cost at the UE for all of them is $\mathcal{C}_{\rm{UE}}=K\stxt{a}$.
In the next section we show an additional advantage of all types of feedback when the feedback is explicitly used to engage re-transmission. We also demonstrate a UE cost reduction by means of the proposed double-threshold feedback. 




\subsection{Integrated feed-forward and feed-back links}
\label{seq:NumB}

In this section we consider a multi-slot transmission where each feed-forward transmission slot is followed with a feedback slot. We consider a system where a user who receives (or infers) a negative feedback is allowed to re-transmit. The user messages are encoded with a Polar code and CRC. At slot $i$, $i=1,2,\cdots$ a set of $K_{\rm{a}}$ active users is allowed to transmit. The set is comprised of re-transmitting users from the previous slot $i-1$, and the new users. The signal $\boldsymbol{y}[i]$ received on the feed-forward link is then split into $\boldsymbol{y}_{\rm{p}}[i]$ which undergoes AD and $\boldsymbol{y}_{\rm{d}}[i]$ for MUD processing. The feed-forward link processing starts with AD that produces the set of preamble signatures, permutations, and scrambling sequences for the detected users. These are then utilized by the MUD algorithm to generate the lists of successfully decoded users ${\mathcal{S}}_{\rm{a}}$ and failed users ${\mathcal{F}}_{\rm{a}}$. These lists are created with the aid of the CRC of the decoded codewords assuming that a failed user has $\text{CRC}\neq 0$. The receiver then uses the preamble sequences of users from sub-sets of the above lists to construct the feedback packet. 

The feedback packet is then broadcasted over the $same$ media and each active user starts processing the received feedback packet. Each active user estimates its channel and the threshold. In case of single-threshold feedback each user match-filters and correlates its (full or partial) signature with the $\underline{\boldsymbol{x}}$ of the feedback packet to infer its status. Active users that detect negative acknowledgment re-transmit in slot $i + 1$. Here, for simplicity, we consider a system where each user can re-transmit only once, while in general the number of possible re-transmissions can be higher. Block fading Rayleigh channel model is considered, where the feed-forward and feedback channel are reciprocal for each slot $i$, but are independent from slot to slot.

Table~\ref{tab:smallAndBigSettings} summarizes feed-forward and feedback settings for two types of systems. System $A$ is designed for smaller number of users, while System $B$ allows for higher number of users, and applies the standard URA parameters used in~\cite{polyanskiy2017perspective} widely adopted in URA literature for comparison purposes.   
\begin{table}
    \centering
    \begin{tabular}{c|c|c}
    \hline 
     & System A & System B \\
     \hline 
     \hline 
     FEC &  \multicolumn{2}{c}{$(511,100)$ Polar} \\ \hline 
     CRC Length &  \multicolumn{2}{c}{$11$} \\ \hline 
        Preamble Length $N_{\rm{p}}$ & $2000$ & $6500-8000$\\ \hline
        Preamble $E_{\rm{b}}/N_{\rm{o}}$ [dB] & $12$ & $15$ \\ \hline 
        Payload Length $N_{\rm{d}}$ & $5500$ & $23500-22000$ \\ \hline 
        Repetition Factor $M$ & $22$ & $89-95$\\\hline
        \hline
        Feedback link $E_{\rm{b}}/N_{\rm{o}}$ [dB] &  \multicolumn{2}{c}{$20$} \\ \hline 
        Targeted $\rm{PUPE}$ & \multicolumn{2}{c}{$0.05$} \\
        \hline 
        Number of re-transmissions & \multicolumn{2}{c}{$1$} \\ \hline 
    \end{tabular}
    \caption{Settings of the feed-forward and feedback links for the two systems under testing.}
    \label{tab:smallAndBigSettings}
\end{table}

Because of re-transmission, the average $E_{\rm{b}}$ is given by 
\begin{eqnarray}
E_{\rm{b}} &=& \frac{K_{\rm{a}}\bar{E}_{\rm{b}} + K_{\rm{a}} \bar{E}_{\rm{b}} P_{\rm{e}}}{K_{\rm{a}}} \nonumber \\
&=& \big( 1 + P_{\rm{e}} \big) \bar{E}_{\rm{b}}, \nonumber  
\end{eqnarray}
consequently, the equivalent $E_{\rm{b}}/N_{\rm{o}}$ accounting for the feedback is given by
\begin{eqnarray}
\frac{E_{\rm{b}}}{N_{\rm{o}}}  = \frac{\bar{E}_{\rm{b}}}{N_{\rm{o}}} + \big( 1 + P_{\rm{e}}\big)_{\rm{dB}},
\end{eqnarray}
where $\bar{E}_{\rm{b}}$ and $P_{\rm{e}}$ are the energy per bit and PUPE per single feed-forward transmission. In this section, we choose the feed-forward $\bar{E}_{\rm{b}}/N_{\rm{o}}$ such that the overall PUPE, accounting for re-transmissions, is $\underline{P}_{\rm{e}} \approx 0.05$.  We then calculate the overall SNR $E_{\rm{b}}/N_{\rm{o}}$ and the average number of new users $\bar{K}\stxt{a}$ that transmit at each time slot. This means that we allow for higher probability of error in the feed-forward link and allow it to be compensated by the feedback to attain the PUPE requirement of the overall performance of the URA system. 

\begin{figure}
    \centering
    \includegraphics[scale = 0.9]{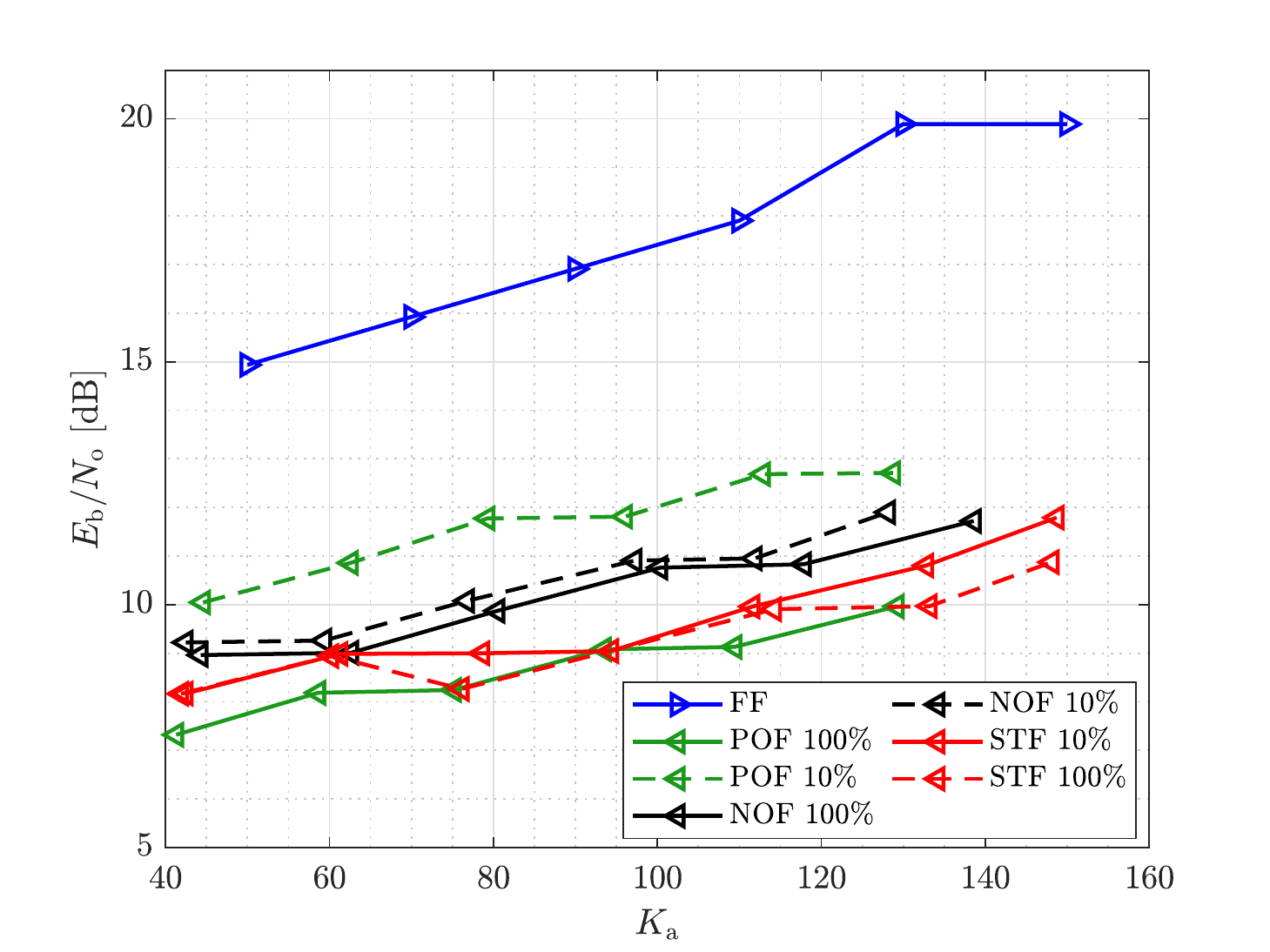}
    \caption{The minimum $E_{\rm{b}}/N_{\rm{o}}$ required to support $\bar{K}\stxt{a}$ new users (unique users) per slot for System A.  
    The targeted probability PUPE after feedback $\underline{P}\stxt{e} \leq 0.05$.}
    \label{fig:polar_integrated_smallSystem}
\end{figure}
Fig.~\ref{fig:polar_integrated_smallSystem}
compares the proposed single-threshold feedback (STF) approach with the positive-only feedback (POF) and negative-only feedback (NOF) approaches for System $A$. 
First of all we can see that use of feedback and even a single re-transmission is far superior in terms of required $E_{\rm{b}}/N_{\rm{o}}$ compared to a feed-forward link without feedback.


Considering feedback which uses full preamble sequences the positive-only approach provides the greatest gain in terms of $E_{\rm{b}}/N_{\rm{o}}$,
compared to the feed-forward link without  feedback. That is because it enables the recovery of both the failed and missed  users.
The performance of the single-threshold feedback approach is very close. It suffers minor performance degradation due to the set $\widehat{F}\stxt{md}$ which infers erroneous feedback. The negative-only approach is inferior since it dismisses the entire set of missed users $F\stxt{md}$. Note that for System A the preamble sequance are relatively long while the numbers of active users are low. Therefore, the correlator used for feedback detection at the user's end is capable of processing all feedback types. This is not the case for partial preamble sequences nor for the System B with a large number of users where the positive-only approach suffers degradation due to the higher number of preamble sequences packed into a shorter feedback message. 


For System A with feedback using shortened preamble sequences the single-threshold approach provides approximately the same performance as in the full sequence case. Similarly only slight increment of the equivalent $E_{\rm{b}}/N_{\rm{o}}$ is observed for the negative-only feedback. For positive-only feedback, however, the reduction of sequence lengths results in a considerable increase of the equivalent $E_{\rm{b}}/N_{\rm{o}}$. This is a direct consequence of increased feedback error caused by the user's sequence correlator. A significant portion of active users with successful decoding status incorrectly infer feedback status and re-transmit. Hence, they simultaneously reduce the number of unique users in the system (new users per slot) and increase the equivalent $E_{\rm{b}}/N_{\rm{o}}$. We can see that for the case of shortened signature the proposed approach is a clear winner, not only in terms of BS cost, but also in terms of performance.

\begin{figure}
    \centering
    \includegraphics[scale = 0.9]{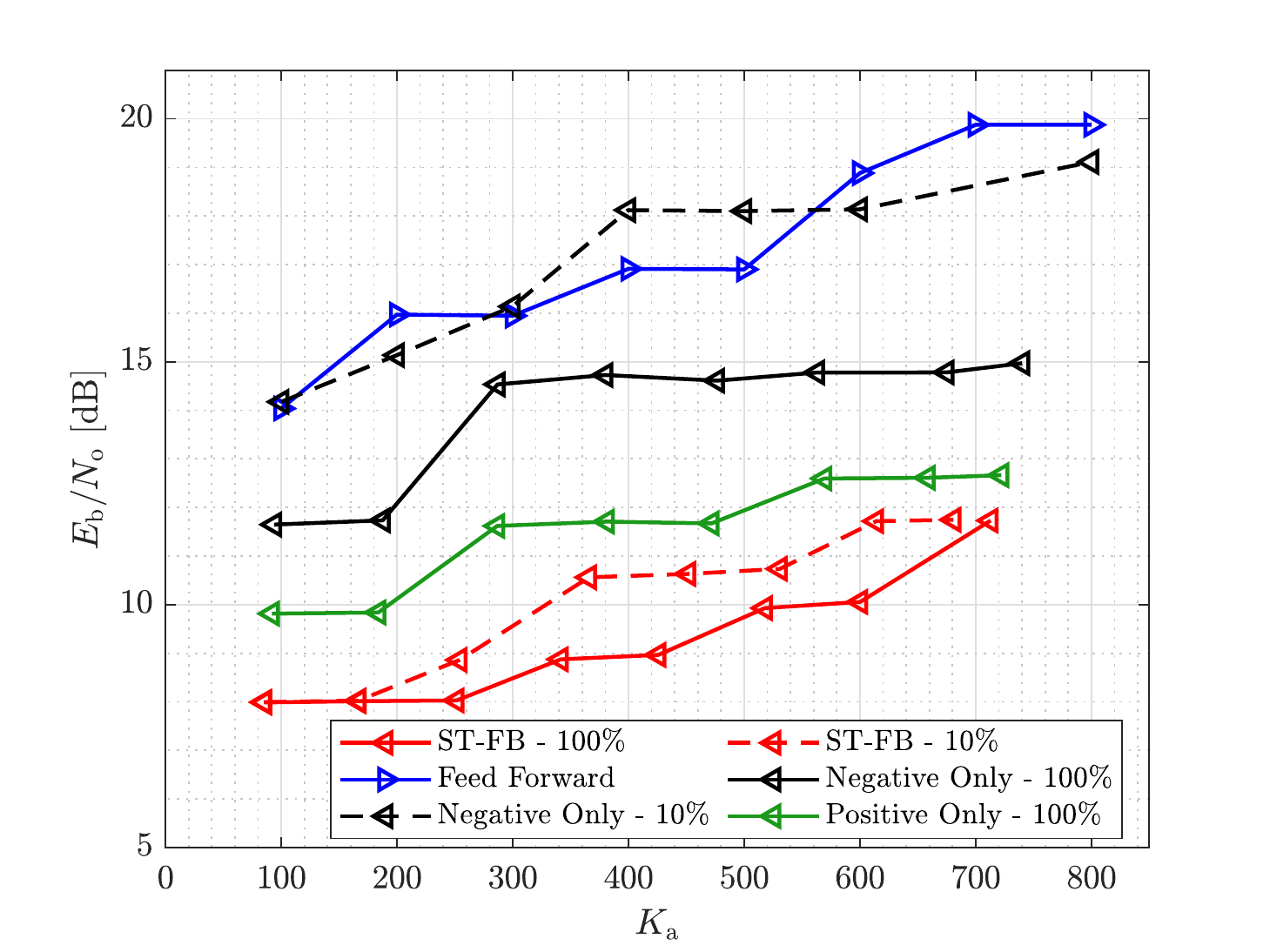}
    \caption{The minimum $E_{\rm{b}}/N_{\rm{o}}$ required to support $\bar{K}\stxt{a}$ new users per slot for System B.  
    The targeted probability PUPE after feedback $\underline{P}\stxt{e} \leq 0.05$.}
    \label{fig:polar_integrated_largeSystem}
\end{figure}

Fig.~\ref{fig:polar_integrated_largeSystem} presents the performance for the case of System $B$ that is able to support much higher number of users than System $A$. The gains in terms of $E_{\rm{b}}/N_{\rm{o}}$ are more pronounced than for System $A$ but, more importantly, the figure presents novel performance-related aspects that are absent in Fig.~\ref{fig:polar_integrated_smallSystem}. Single-threshold feedback approach provides a significant gain over the positive-only feedback even for the full sequence case. Evidently, much lower preamble sequence density of the proposed approach compared to the positive-only feedback permits the user's correlator to have lower miss detection rates and better performance. 
All feedback systems deliver appreciable gains over the single feed-forward link. 
Note that the number of unique users in these systems is slightly reduced compared to the feed-forward link only (each simulation point is slightly pulled to the left). Yet, any system with feedback with full-length sequences undergoes the overall throughput reduction by a factor of $\frac{N_{\rm{p}}}{N_{\rm{p}} + N_{\rm{d}}}$ which is almost $26\%$ assuming that both feed-forward and feedback links share the same channel resource. 

The dashed curves show the system's performance with feedback using shortened preamble sequences. 
The sequences that are shortened to the $10\%$ of the original length lead to the very minor $2.6\%$ reduction in the throughput compared to the feed-forward only link. 
The performance of the negative-only feedback for System B with partial preamble sequences is severely degraded compared to the full-sequence case, while positive-only approach for this case is completely infeasible for $\bar{K}\stxt{a}\geq 100$. On the contrary, the proposed single-threshold feedback approach suffers almost no performance degradation. As we already noted, this is cased by the significant reduction of the BS cost of the feedback (number of preamble sequences packed into the feedback signal), which, in turn, reduces the error of the feedback processing at the user's side.

\subsection{Double Threshold Feedback}

\begin{figure}
    \centering
    \begin{tabular}{cc}
    \includegraphics[scale = 0.555]{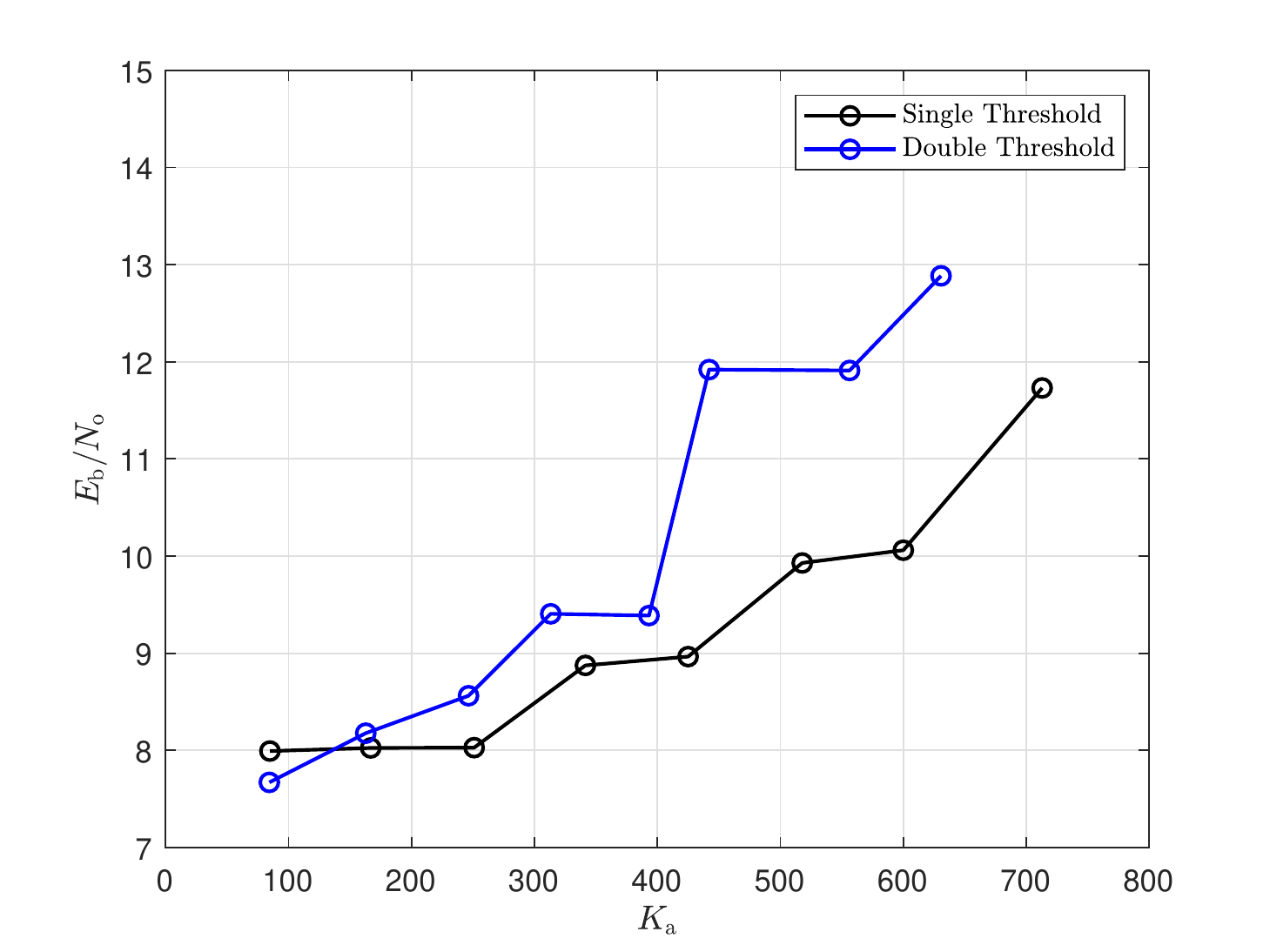} &  \includegraphics[scale=0.555]{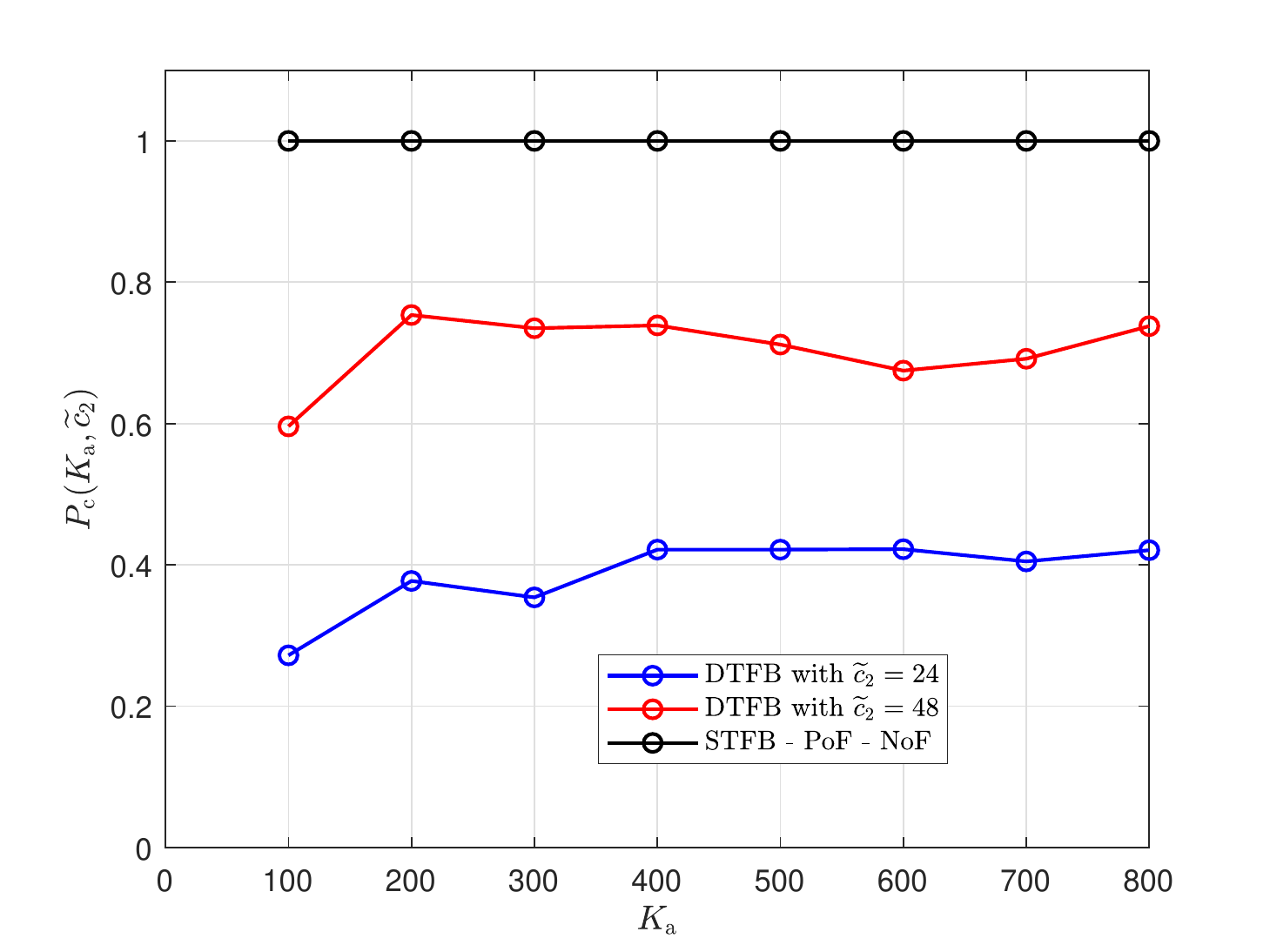} \\
    $(a)$ & $(b)$ 
\end{tabular}
    \caption{$(a)$ The minimum $E_{\rm{b}}/N_{\rm{o}}$ required for PUPE $\leq 0.05$ for System B with single and double threshold feedback. $(b)$ The probability that a user reaches the final stage of processing of the feedback signal. The lower threshold value is $\widetilde{c}_1 = 12$ for the double-threshold system while the optimal $\widetilde{c}$ is selected for the single-threshold system.} 
    \label{fig:doubleThreshold}
\end{figure}
Fig.~\ref{fig:doubleThreshold} presents the multi-slot transmission results of the single and double-threshold feedback schemes for the System B. Because of the fact the all users below the lower threshold $\widetilde{c}_1$ re-transmit regardless of their status, the number of re-transmitting users in the double-threshold feedback system is higher than for the single-threshold method. This translates into an increase of the equivalent required $E_{\rm{b}}/N_{\rm{o}}$ compared to the single-threshold approach. As shown in Fig.~\ref{fig:doubleThreshold}~$(a)$, the gap widens as the number of users increases. 
Nevertheless the double-threshold approach provides a significant $E_{\rm{b}}/N_{\rm{o}}$ gain over the feed-forward link.

Fig.~\ref{fig:doubleThreshold}~$(b)$ demonstrates the UE cost advantage of the double-threshold approach. The probability $P_{\rm{c}}=\mathcal{C}_{\rm{UE}}^{\rm{d}}/\bar{K}\stxt{a}$ defines  the frequency of the utilization of the user's correlator used for feedback reception. All active users in single-threshold, positive, and negative-only feedback approaches always have to use their correalator. Hence, $P_{\rm{c}} = 1$ for the single-threshold approach. For the double-threshold approach, however, $P_{\rm{c}}$ and the respective UE cost $\mathcal{C}_{\rm{UE}}^{\rm{d}}$ are significantly lower. 
\section{Conclusions and Future Work}
\label{sec:Con}
In this paper we focus on a systematic approach to feedback design for URA systems. Our proposed  feedback format allows the URA users to acquire the knowledge about the status of the packet transmission in a low-complexity fashion, suitable for low-power miniature UEs. We have also shown that the typical positive-only and negative-only feedback approaches are not suitable for URA systems with a large numbers of active users. We have demonstrated that the feedback leads to the improvement of the system performance in terms of the PUPE and the SNR required to support various numbers of active users. Our proposed  single-threshold feedback approach reduces the number of preambles included into the feedback signal and enables a simple correlation detection receiver at the UE side. Our second proposed algorithm, the double-threshold approach, allows most of the users to skip the processing of the main part of the feedback signal and save the power even further. 



%





\ifCLASSOPTIONcaptionsoff
  \newpage
\fi



%


\bibliographystyle{IEEEtran}
\bibliography{ms}
%








\end{document}